\documentclass[a4,aps,prd,10pt,twocolumn,notitlepage,superscriptaddress]{revtex4-2}
\usepackage{amsmath,amssymb,amsfonts}
\usepackage{bbm,bm}
\usepackage[utf8]{inputenc}
\usepackage{amsmath,amsfonts,amssymb,graphicx,subfigure}
\usepackage{slashed,color}
\usepackage{multirow}
\usepackage{lineno}
\usepackage{verbatim}
\usepackage[svgnames]{xcolor}
\usepackage{hyperref}

\usepackage{hyperref}
\hypersetup{colorlinks=true, linkcolor=blue, citecolor=ForestGreen,
filecolor=magenta, urlcolor=MidnightBlue,
  pdftitle={Basu and Ruiz [arXiv:]},
  pdfauthor={Basu and Ruiz},
  pdfsubject={Subject},pdfcreator={Creator},pdfproducer={Producer},
  pdfkeywords={Helicity Polarization}{Hadron Colliders}{V+jets}{High Orders}
}
\graphicspath{{Figs/}}

\def\GeV{{\rm\ GeV}}
\def\TeV{{\rm\ TeV}}
\def\jets{{\rm jets}}

\definecolor{magenta}{HTML}{FF00FF}
\definecolor{cornflowerblue}{HTML}{6495ED}
\definecolor{turquoise}{HTML}{40E0D0}

\definecolor{darkgreen}{rgb}{0.0, 0.2, 0.13}
\definecolor{darkmagenta}{rgb}{0.55, 0.0, 0.55}
\definecolor{amber}{rgb}{1.0, 0.6, 0.0}

\newcommand{\orcid}[1]{\,\href{https://orcid.org/#1}{\includegraphics[width=9pt]{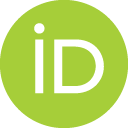}}}
\newcommand{\orcidTB}{0009-0003-1414-0282} 
\newcommand{\orcidRR}{0000-0002-3316-2175} 

\begin{document}
\rightline{IFJPAN-IV-2026-13, COMETA-2026-09} 

\title{Polarization interference in exclusive
\texorpdfstring{$V$+jets}{V+jets} 
at all orders in 
\texorpdfstring{$\alpha_s$}{QCD}}

\author{Trina Basu\orcid{\orcidTB}}
\email{trina.basu@ifj.edu.pl}

\author{Richard Ruiz\ \orcid{\orcidRR}}
\email{rruiz@ifj.edu.pl}

\affiliation{Institute of Nuclear Physics -- Polish Academy of Sciences {\rm (IFJ PAN)}, ul. Radzikowskiego, Krak{\'o}w, 31-342, Poland}


\begin{abstract}
Using new methods for computing helicity amplitudes 
with intermediate helicity-polarized gauge bosons,
we revisit the transverse-longitudinal 
polarization interference 
in the $pp\to V+{\rm jets}$ process 
for $V=\gamma^*,Z^{(*)},W^{(*)}$ decaying to massless leptons.
At each order of the strong coupling constant $\alpha_s$
and remaining exclusive with respect to jet kinematics, 
we show that the polarization interference 
in $\gamma^*\to\ell^+\ell^-$  vanishes 
after phase-space integration over the kinematics of $\ell^\pm$,
thereby extending well-known results for the inclusive process.
Due to parity violation, cancellations 
are softened for the $W$ and $Z$ bosons.
We give a simple formula to account for fiducial cuts.
We comment on the implications for multiboson processes,
and the applicability of our results 
to chiral gauge bosons in new physics scenarios
and
to polarization measurements of weak bosons 
in heavy ion collisions.
\end{abstract}

\date{\today}

\maketitle

\textbf{Introduction}
The universe's transition from an ensemble of massless states
to today's collection of massive particles 
remains one of the least understood 
periods of the early universe.
With the discovery~\cite{ATLAS:2012yve,CMS:2012qbp} 
and classification~\cite{ATLAS:2022vkf,CMS:2022dwd} 
of the Higgs boson,
the Standard Model of particle physics
now gives a prediction for 
the electroweak (EW) phase 
transition~\cite{Morrissey:2012db,DiMicco:2019ngk}.
However, it is far from established 
whether nature realizes this transition 
entirely through the minimal Higgs scenario,
or if new interactions modify the Standard Model picture~\cite{Profumo:2007wc,Chala:2018ari,Ramsey-Musolf:2019lsf,Athron:2023xlk}.

Precise measurements of Higgs production 
and multiboson processes 
at the Large Hadron Collider (LHC) 
(and proposed successors)
are therefore essential as they probe 
EW symmetry breaking and are sensitive to new 
phenomena~\cite{EuropeanStrategyforParticlePhysicsPreparatoryGroup:2019qin,
P5:2023wyd,deBlas:2025gyz}.
Measuring the transverse $(\lambda=\pm1)$ 
and longitudinal $(\lambda=0)$
helicity polarizations of the $W$ and $Z$ gauge bosons 
is particularly interesting as this constitutes 
a rare probe of the subtle but structural cancellations that 
render spontaneously broken non-Abelian gauge theories 
renormalizable and unitary~\cite{tHooft:1971qjg,
Becchi:1974md,Becchi:1974xu,Becchi:1975nq,
Tyutin:1975qk}.

\begin{figure*}[t!]
  \includegraphics[width=.75\textwidth]{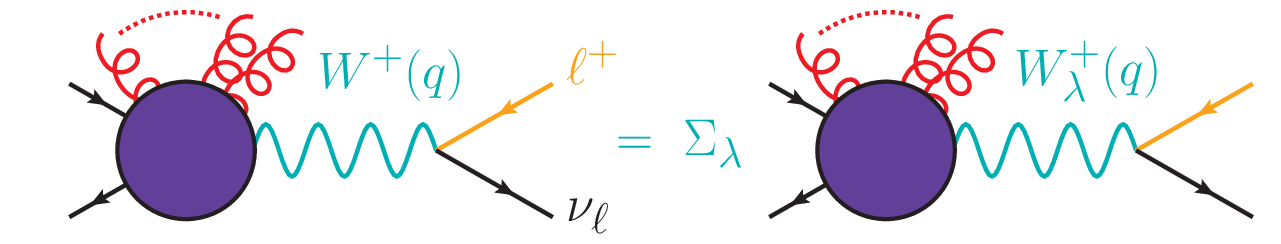}
  \caption{Illustration of the relationship between the 
  $pp\to\ell^+\nu_\ell+\jets$ process 
  for (L) unpolarized $W^+$ and (R) polarized $W_\lambda^+$.
  }
\label{diagram_wJets_PolarMEdef}
\end{figure*}

In recent years, the polarization program 
at the LHC 
has relied on predictions for scattering rates
using the so-called ``polarized propagator''
formalism~\cite{Ballestrero:2017bxn,Ballestrero:2019qoy}
and its extensions~\cite{BuarqueFranzosi:2019boy,
Denner:2020bcz,Hoppe:2023uux,Javurkova:2024bwa}.
The framework models the propagation 
of intermediate weak bosons or heavy quarks 
as superpositions of their helicity eigenstates,
$\vert \psi_{\rm unpol}\rangle\langle \psi_{\rm unpol}\vert = 
\sum_{\lambda}\vert \psi_\lambda\rangle\langle \psi_{\lambda}\vert$.
As illustrated in 
Fig.~\ref{diagram_wJets_PolarMEdef}
for $W+\jets$,
one can then conceptualize the amplitude 
for the production of an unpolarized state 
$(\mathcal{M}_{\rm unpol})$
as the coherent sum over amplitudes for polarized states 
$(\mathcal{M}_\lambda)$.

At the squared amplitude level, 
the unpolarized process 
is then given by squared polarized contributions 
and the
interference among different helicity polarizations,
\begin{align}
\vert \mathcal{M}_{\rm unpol}\vert^2 = \sum_\lambda \vert\mathcal{M}_\lambda\vert^2 
+ \sum_{\lambda\neq\lambda'}\mathcal{M}_\lambda^*\mathcal{M}_{\lambda'}\ .
\end{align}
Here, 
$\mathcal{M}_{\rm unpol}$ and $\mathcal{M}_{\lambda}$
are collections of subamplitudes / diagrams
describing (near) resonant production of weak boson $V$
such that $\mathcal{M}_{\rm unpol}=\sum_\lambda \mathcal{M}_\lambda$ is gauge invariant.

In some sense, this organization is a realization 
of BRST invariance~\cite{tHooft:1971qjg,
Becchi:1974md,Becchi:1974xu,Becchi:1975nq,Tyutin:1975qk}
because it puts helicity-polarized gauge bosons $V_\lambda$
(and their gauge-fixed propagators)
on the same level as Goldstone bosons and other
degrees of freedom from the gauge-fixing sector~\cite{Javurkova:2024bwa,Basu:2025zds}.
Importantly, this formalism has been incredibly successful 
in describing polarization data from 
the LHC~\cite{CMS:2011kaj,CMS:2020etf,CMS:2021icx,
ATLAS:2022oge,ATLAS:2023zrv,ATLAS:2024qbd},
can be implemented systematically in precision 
calculations~\cite{Baglio:2018rcu,Baglio:2019nmc,
Denner:2020bcz,Denner:2020eck,Denner:2021csi,Denner:2022riz,
Le:2022lrp,Le:2022ppa,Carrivale:2025mjy,Pelliccioli:2025com},
and even automated in general-purpose Monte Carlo event 
generators~\cite{BuarqueFranzosi:2019boy,
Hoppe:2023uux,Javurkova:2024bwa}.

In this work, we revisit predictions for the
production and decay of polarized EW gauge bosons 
$V_\lambda$
in the $pp\to V+\jets$ process.
At each order in the strong coupling constant $\alpha_s$,
we derive kinematical conditions 
under which the longitudinal-transverse 
polarization interference in $V+\jets$ 
vanishes or is at the percent level.

Departing from past studies~\cite{Lam:1978pu,
Bern:2011ie,Stirling:2012zt,Belyaev:2013nla,Pellen:2021vpi},
we work entirely in the polarized propagator formalism,
without imposing the narrow width or pole approximations,
and compute polarization interference directly.
For the special case $V=\gamma^*$, 
we recover the well-known results of Lam \& Tung~\cite{Lam:1978pu}.
Unlike Ref.~\cite{Lam:1978pu},
which used structure functions 
to model hadronic activity
and formally holds only for inclusive hadronic processes,
we remain exclusive with respect to hadronic activity.
For the weak bosons,
we show that (non-)vanishing polarization interference 
is inherently tied to parity violation in the weak sector,
explaining observations reported in  
Refs.~\cite{Denner:2020bcz,Denner:2020eck}
and demonstrating broader applicability of 
Refs.~\cite{Azatov:2016sqh,Panico:2017frx}.

To carry out our work, 
we exploit newly identified expressions 
for polarized gauge boson propagators, 
reported in a companion study~\cite{Basu:2025zds}.
These identities simplify the organization and evaluation 
of polarized amplitudes $\mathcal{M}_\lambda$
and make power counting of mass-over-energy factors manifest 
in covariant gauges.

\textbf{Setup:}
Our goal is to build an expression 
for the transverse-longitudinal polarization interference 
in the $pp\to V+\jets$ process, 
denoted by $\mathcal{I}_{\rm pol}^{V+\jets}$,
at the totally differential level and at an arbitrary order of $\alpha_s$.
To do this, we consider the $2\to (n+2)$ process
\begin{align}
    i(p_i)+j(p_j) &\to \sum_{m=1}^n k_m(k_m) + V^{(*)}_\lambda(q)\  
    \nonumber\\
     &\to \sum_{m=1}^n k_m(k_m) + f_h(p_f) 
     + \overline{f'_{h'}}(p_{f'}) ,
\label{eq:def_proc_wjets}    
\end{align}
for partons $i,j,k_m\in\{q,\overline{q'},g\}$
and leptons $f,\overline{f'}\in\{\ell,\nu_\ell\}$.
Momentum assignments are given in parentheses,
with the momentum of the intermediate $V_\lambda$ given by
$q = p_f+p_{f'} = p_i+p_j - (\sum_{m=1}^n k_m)$.
We denote the helicities of $V$, $f$, and $\overline{f'}$ by 
the subscripts $\lambda$, $h$, and $h'$.

Crucial to our work are the following assumptions:

(i) We work at lowest order in the EW theory 
and with massless leptons.
This implies that all diagrams for Eq.~\eqref{eq:def_proc_wjets} contain 
$s$-channel $V_\lambda$ exchanges, 
as in Fig.~\ref{diagram_wJets_PolarMEdef}.
This also guarantees that the outgoing lepton current
for arbitrary chiral couplings $g_L$ and $g_R$,
\begin{align}
J_{h,h'}^\mu\ =\ \overline{u}(p_f,h) \gamma^\mu
\left(g_LP_L + g_RP_R\right) 
v(p_{f'},h')\ ,
\label{eq:def_lep_current}
\end{align}
 where $P_{L/R}=\frac{1}{2}(1\mp \gamma^5)$, 
 is a conserved current
\begin{align}
\label{eq:current_conservation_lep}
    q_\mu\cdot J_{h,h'}^\mu\ 
    =\ (p_f+p_{f'})_\mu \cdot J_{h,h'}^\mu\ 
    =\ 0\ .
\end{align}

(ii) We work in massless quantum chromodynamics (QCD).
This precludes the production of $V$ 
from the decays of heavy quarks
and guarantees that 
for a fixed order in $\alpha_s$ 
the sum of all $2\to n$ ``incoming'' currents
\begin{align}
\label{eq:def_qcd_current}
 G^\nu_{in}\ &=\ \sum_{\sigma\in\{i,j,k_m\ {\rm configurations}\}}\ G^\nu_{\sigma}
 \\
 G^\nu_{\sigma}\ &\equiv\ \overline{v}\left[\dots \delta^{AB}\left(g_LP_L + g_RP_R\right)\gamma^\nu\ \cdots\right]u\ ,
\end{align} 
is also a conserved current with respect to $q^\mu$ and $V_\lambda(q)$,
\begin{align}
\label{eq:current_conservation_jets}
    q_\nu\cdot G^\nu_{in}\ =\  
    E_VG^{\nu=0}_{in} - \vert\vec{q}\vert\hat{q}\cdot \vec{G}_{in}\ =\
    0\ .
\end{align}

In our organization, 
$G^\nu_{\sigma}$ is any $(n+2)$-point 
Green's function with $(n+2)$ external quarks and gluons,
all spinor indices and gluon vertices contracted,
but features one uncontracted Lorentz index
that corresponds to the $V$-quark-antiquark vertex.
$G^\nu_{in}$ is then the sum over all possible $G^\nu_{\sigma}$
at a particular order in $\alpha_s$.

Current conservation for $G^\nu_{in}$ is a non-trivial consequence 
of $V$ being colorless and QCD being unbroken.
Since $V$ is a color singlet, the sum over all $i,j,k_m$
configurations must lead to a contraction of all color indices.
(This is more obvious for the process $e^+e^-\to\jets$.)
In unbroken, non-abelian theories, 
\textit{covariant} current conservation $\mathcal{D}_\nu G^\nu_{in}=0$
holds when $G^\nu_{in}$ is gauge invariant with respect to that gauge symmetry~\cite{Bohm:2001yx}.
The stronger condition of Eq.~\eqref{eq:current_conservation_jets}
follows because $V$ is colorless.

Like Eq.~\eqref{eq:current_conservation_lep},
Eq.~\eqref{eq:current_conservation_jets}
is essentially the momentum-space representation 
of the Slavnov-Taylor identity 
$\partial_\nu G^\nu_{in}=0$ in coordinate space.
We stress that our results follow from these relationships
applied to polarized propagators,
rather than the evaluation of any individual diagram.

(iii) While we remain exclusive 
with respect to the kinematics of the $n$-jet system, 
we assume that the incoming partonic center-of-mass scale 
$\sqrt{\hat{s}}=\sqrt{2p_i\cdot p_j}$
and the hard scattering scale $Q = \sqrt{q^2}$ 
are large enough 
to regulate the $\gamma^*\to f\overline{f}$ splitting 
in Eq.~\eqref{eq:def_proc_wjets}.
Importantly, on-shell kinematics 
are not required to define helicity polarization~\cite{Lam:1978pu}
or to use polarized propagators~\cite{Javurkova:2024bwa,Basu:2025zds}.

\textbf{Unpolarized and Polarized Amplitudes:}
To construct polarized amplitudes, 
we follow Refs.~\cite{Ballestrero:2017bxn,Ballestrero:2019qoy}
and use completeness 
to decompose the propagator for an unpolarized EW gauge boson 
$\Pi_{\mu\nu}^V(q)$
into the outer product of polarization vectors.
The individual propagators for polarized gauge bosons 
$\Pi_{\mu\nu}^V(q,\lambda)$ are then given by this outer product.
In the general covariant gauge $(R_\xi)$, 
\begin{align}
\label{eq:prop_unpol_rx}
    \Pi_{\mu\nu}^V(q) &=
  -i\left[g_{\mu\nu} + \cfrac{(\xi-1)q_\mu q_\nu}{D_V(q^2,\xi)}\right]\ 
  D_V^{-1}(q^2)
  \\
  &= \sum_{\lambda=\pm1,0,S}\  \Pi_{\mu\nu}^V(q,\lambda)\ ,
\label{eq:completeness}
\\
\Pi_{\mu\nu}^V(q,\lambda) &=
    i \eta_\lambda\ \varepsilon_\mu(q,\lambda)\varepsilon^*_\nu(q,\lambda)\ 
    D_V^{-1}(q^2)\ .
\label{eq:prop_pol_rx}
\end{align}

Here, $\varepsilon_\mu(q,\lambda)$ are 
the usual polarization vectors in the helicity basis
for spin-1 states with virtuality $\sqrt{q^2}$. 
For these, we adopt the \texttt{HELAS} 
convention~\cite{Hagiwara:1985yu,Murayama:1992gi}.
$D_V(q^2)=q^2-M_V^2+iM_V\Gamma_V$ and 
$D_V(q^2,\xi)=q^2-\xi M_V^2+i\xi M_V\Gamma_V$ 
are the pole structures of propagators in the $R_\xi$ gauge,
with $\xi$ being the gauge-fixing parameter.
For the phase factors $\eta_\lambda$,
 we adopt the convention of Refs.~\cite{Dreiner:2008tw,Basu:2025zds} and set
$\eta_{\lambda=S}=-(\eta_{\lambda=0,\pm1})=-1$.
For photons, $M_\gamma,\Gamma_\gamma\to0$.

The sum in Eq.~\eqref{eq:completeness}
runs over the transverse $\lambda=\pm1$ states of $V(q)$,
its longitudinal state $\lambda=0$, 
and a ``scalar'' state $\lambda=S$.
The last two are sensitive to gauge fixing.
This organization allows us 
to follow the philosophy 
of Refs.~\cite{tHooft:1971qjg,
Becchi:1974md,Becchi:1974xu,Becchi:1975nq,Tyutin:1975qk,Becchi:2014lsa}, 
which argue that longitudinally polarized weak bosons 
should not be 
treated in isolation but in conjunction with 
gauge-fixing degrees of freedom.
Throughout this work, 
we sum over the transverse polarizations, 
denoting the combination as $\lambda=T$.

A technical novelty of our work is the observation 
that the propagators in Eq.~\eqref{eq:prop_pol_rx}
have simple decompositions that 
separate transverse $(\Phi_{\mu\nu})$ 
and forward-backward $(\Theta_{\mu\nu})$ 
propagation of $V_\lambda$.
These are given by
\begin{subequations}
\label{eq:prop_pol_rx_phi_theta}
\begin{align}
\Pi_{\mu\nu}^V(q,\lambda=T) &= i\ \Phi_{\mu\nu}\ D_V^{-1}(q^2)   
\\
\Pi_{\mu\nu}^V(q,\lambda=0) &= i\ \left[\Theta_{\mu\nu}+ \frac{q_\mu q_\nu}{q^2}\right]\ D_V^{-1}(q^2)   
\\
\Pi_{\mu\nu}^V(q,\lambda=S) &= 
\frac{-i}{D_V(q^2)}  
\left[
\frac{q_\mu q_\nu}{q^2}
+
\frac{(\xi-1)q_\mu q_\nu}{D_V(q^2,\xi)}
\right] ,
\end{align}    
\end{subequations}
where the Lorentz-covariant tensor structures are
\begin{subequations}
\label{eq:prop_pol_tensors}
\begin{align}
 \Phi_{\mu\nu}\ &=\ \hat{q}_{\perp\mu}\hat{q}_{\perp\nu}\ +\ \hat{q}_{T\perp\mu}\hat{q}_{T\perp\nu}\ ,
 \\ 
 \Theta_{\mu\nu}\ &=\ 
    \frac{(n\cdot q)}{(n\cdot q)^2 - q^2 n^2}
    \Big[-n_\mu q_\nu - q_\mu n_\nu 
    \nonumber\\
    &\ 
    +\ \frac{q_\mu q_\nu n^2}{(n\cdot q)}\
    +\ \frac{n_\mu n_\nu q^2}{(n\cdot q)}\Big]\ ,
\end{align}
\end{subequations}
and completeness is preserved via the exact relationship
\begin{align}
 -g_{\mu\nu}\ &=\ \Phi_{\mu\nu}\ +\ \Theta_{\mu\nu}\  .
\end{align}

Here, 
$q^\mu=(E_V, q_x, q_y, q_z)$ is $V$'s momentum 
in an arbitrary frame, 
with $\hat{q} = \vec{q}/\vert\vec{q}\vert$
being the direction of its 3-momentum.
$n^\mu=(1, -\hat{q})$ is a light-like vector ($n^2=0$)
with $n\cdot q = E_V+\vert \vec{q}\vert$.
$\hat{q}_{\perp}^\mu = 
(1/q_T\vert\vec{q}\vert)(0, q_x q_z, q_y q_z, -q_T^2 )$
and
$\hat{q}_{T\perp}^\mu = 
(1/q_T)(0, - q_y, q_x, 0 )$ 
are space-like vectors 
($\hat{q}_{\perp}^2,\ \hat{q}_{T\perp}^2=-1$),
are mutually orthogonal, 
and are orthogonal to $q^\mu$ and $n^\mu$.

The expressions in Eq.~\eqref{eq:prop_pol_rx_phi_theta}
are notable as we are working with massive gauge bosons 
in a covariant gauge.
The transverse tensor $\Phi_{\mu\nu}$ has long been 
identified in propagators for photons and 
gluons~\cite{Beenakker:1993yr,Aivazis:1993kh,Bohm:2001yx},
while the forward-backward tensor $\Theta_{\mu\nu}$ 
is usually associated with propagators 
and power counting in axial gauges~\cite{Sterman:1978bi,
Libby:1978bx,Kunszt:1987tk,Sterman:1995fz,Dams:2004vi}.
In Ref.~\cite{Dittmaier:2025htf}, 
structures similar to $\Phi_{\mu\nu}$ and $\Theta_{\mu\nu}$ 
were identified but restricted to on-shell momentum $(q^2=M_V^2)$.

Using these expressions, 
the unpolarized $(\mathcal{M}_{\rm unpol})$ 
and polarized $(\mathcal{M}_{\lambda})$
amplitudes for $V+\jets$ 
in terms of incoming and outgoing current 
$G_{in}^\nu$ and $J_{\lambda\lambda'}^\mu$  
are 
\begin{subequations}
\label{eq:inter_rx_matrix}
\begin{align}
    -i\mathcal{M}_{\rm unpol} =&\ 
    J_{h,h'}^\mu\ 
    i 
    \left[-g_{\mu\nu}
    -\frac{(\xi-1)q_\mu q_\nu}{D_V(q^2,\xi)}
    \right] 
    G_{in}^\nu\ D_V^{-1}(q^2)
    \nonumber\\
    =&\ J_{h,h'}^\mu\ 
    i \left[-g_{\mu\nu}    \right]
    G_{in}^\nu\ D_V^{-1}(q^2),
    \\
    -i\mathcal{M}_{\lambda=T} =&\ 
    J_{h,h'}^\mu\ 
    i \left[\Phi_{\mu\nu}    \right]
    G_{in}^\nu\ D_V^{-1}(q^2)\ ,
\\
    -i\mathcal{M}_{\lambda=0}=&\ 
    J_{h,h'}^\mu\ i
    \Big[\Theta_{\mu\nu}+ \frac{q_\mu q_\nu}{q^2}
    \Big]
    G_{in}^\nu\ D_V^{-1}(q^2)\ 
    \nonumber\\
    =&\ 
    J_{h,h'}^\mu\ i
    \left[n_\mu n_\nu\frac{q^2}{(n\cdot q)^2}\right]
    G_{in}^\nu\ D_V^{-1}(q^2) 
    ,
\\
    -i\mathcal{M}_{\lambda=S}=&\ 
    J_{h,h'}^\mu i 
    \left[
    -\frac{q_\mu q_\nu}{q^2}
    -\frac{(\xi-1)q_\mu q_\nu}{D_V(q^2,\xi)}
    \right]
    G_{in}^\nu 
    D_V^{-1}(q^2)
    \nonumber\\
    =&\ 0\  .
\end{align}
\end{subequations}
Due to current conservation 
the $\lambda=S$ amplitude vanishes and most other amplitudes simplify.
There are no Goldstone amplitudes 
nor are there EW ghost amplitudes
since all external particles are massless 
and we are working at lowest order in the EW theory.

\textbf{Polarization Interference:}
Since there are no scalar or Goldstone contributions,
the totally unintegrated polarization interference 
$(\mathcal{I}_{\rm pol}^{V+\jets})$ is determined  
by the product of the transverse and longitudinal amplitudes.
Using Eq.~\eqref{eq:inter_rx_matrix}, we have
\begin{align}
  \mathcal{I}_{\rm pol}^{V+\jets}(h,h')\ &=\ 
  2\Re\left[\mathcal{M}_{\lambda=0}^*\mathcal{M}_{\lambda=T}\right]\ ,
  \quad\text{where}
  \\
\mathcal{M}_{\lambda=0}^*\mathcal{M}_{\lambda=T} &= 
    \frac{q^2}{(n\cdot q)^2}
    \frac{(J_{h,h'}\cdot n)^*(n\cdot G_{in})^*}{\vert D_V(q^2)\vert^{2}}\ \times\
    \nonumber\\
    \big[(J_{h,h'}\cdot \hat{q}_{\perp})
    (\hat{q}_{\perp}&\cdot G_{in})
    + 
    (J_{h,h'}\cdot \hat{q}_{T\perp})
    (\hat{q}_{T\perp}\cdot G_{in})\big]\ .
\end{align}

The reference vector $n^\mu$ is not Lorentz covariant.
However, its contractions with Lorentz-covariant quantities 
are invariant under rotations~\cite{Hagiwara:2020tbx,Chen:2022gxv}, 
such as the set of rotations $R^\mu_\nu(\hat{i};\theta)$ 
that align $q^\mu$ along the $\hat{z}$ axis,
\begin{align}
    \tilde{q}^\mu\ &=\ [R^{-1}(y,\theta_V)\cdot R^{-1}(z,\phi_V) \cdot q]^\mu 
    \nonumber\\ 
    &=\ (E_V, 0, 0, \vert \vec{q}\vert) .
\end{align}

Applying these rotations 
to $n^\mu$, $\hat{q}_{\perp}^\mu$, 
and $\hat{q}_{T\perp}^\mu$ gives
\begin{align}
\tilde{n}^\mu\ =\ (1,0,0,-1)\ ,\
\tilde{q}_\perp^\mu\ &=\ (0,1,0,0)\ ,\ 
\nonumber\\
\tilde{q}_{T\perp}^\mu\ &=\ (0,0,1,0)\ ,
\end{align}
showing that the $\Pi_{\mu\nu}^V(q,\lambda)$
naturally act as orthogonal projection operators 
relative to direction of $V$'s motion
in the original frame that defines  $q^\mu$ and $\lambda$.
For example: defining $\tilde{G}_{in}^\nu \equiv 
[R^{-1}(y,\theta_V)\cdot R^{-1}(z,\phi_V) \cdot G_{in}]^\nu$
and applying current conservation  [Eq.~\eqref{eq:current_conservation_jets}] 
gives 
\begin{align}
    (n\cdot G_{in})\ &=\ (\tilde{n}\cdot\tilde{G}_{in})\ 
    =\ \frac{E_V+\vert\vec{q}\vert}{E_V}\ \tilde{G}_{in}^3\ .
\end{align}

To build the analogous quantity $\tilde{J}_{h,h'}^\mu \equiv 
[R^{-1}(y,\theta_V)\cdot R^{-1}(z,\phi_V) \cdot J_{h,h'}]^\mu$,
we evaluate the lepton current $J_{h,h'}^\mu$
for the two non-vanishing helicity configurations 
of $f$ and $\overline{f'}$ in the $(f\overline{f'})$ frame using the momenta
\begin{align}
    \bar{q}^\mu &
    = \bar{p}_f^\mu+\bar{p}_{f'}^\mu
    = (\sqrt{q^2},\vec{0})
    \quad,\quad
    \bar{E}_f = \frac{\sqrt{q^2}}{2}\ ,
    \\
    \bar{p}_f^\mu &= \bar{E}_f (1,\sin\bar\theta_f\cos\bar\phi_f,\sin\bar\theta_f\sin\bar\phi_f,\cos\bar\theta_f)\ ,\
\end{align}
and boost the resultants along the $\hat{z}$-axis 
with the Lorentz factor $\gamma_V = E_V/\sqrt{q^2}$.

Inserting everything into 
$\mathcal{I}_{\rm pol}^{V+\jets}(h,h')$,
we get 
\begin{subequations}
\label{eq:polint_vjets}
\begin{align}
&\mathcal{I}_{\rm pol}^{V+\jets}(h,h')\ =\ 
  2\Re\left[\sum_{k=1,2}\ I_{h,h'}^k\right]\ ,
\end{align}
\begin{align}
I^{1}_{LR} &= 
\vert g_L^f\vert^2 
\tilde{r}_{in}^{k=1}
\sin\bar\theta_f\left(\cos\theta_f\cos\bar\phi_f-i\sin\bar\phi_f\right)\ ,
\\
I^{2}_{LR} &= 
\vert g_L^f\vert^2 
\tilde{r}_{in}^{k=2}
\sin\bar\theta_f\left(\cos\bar\theta_f\sin\bar\phi_f+i\cos\bar\phi_f\right)\ ,
\\
I^{1}_{RL} &= 
\vert g_R^f\vert^2 
\tilde{r}_{in}^{k=1} 
\sin\bar\theta_f\left(\cos\bar\theta_f\cos\bar\phi_f+i\sin\bar\phi_f\right)\ ,
\\
I^{2}_{RL} &= 
\vert g_R^f\vert^2 
\tilde{r}_{in}^{k=2} 
\sin\bar\theta_f\left(\cos\bar\theta_f\sin\bar\phi_f-i\cos\bar\phi_f\right)\ ,
\\
& \tilde{r}_{in}^{k} = 
-q^2 \frac{\sqrt{q^2}}{E_V}
\tilde{G}_{in}^{\nu=k}
(\tilde{G}_{in}^{\nu=3})^*
\vert D_V(q^2)\vert^{-2}\ .
\label{eq:def_polint_rk}
\end{align}
\end{subequations}

Now, the volume element for an $(n+2)$-body phase space 
can always be decomposed into a convolution of volume elements 
for an $n$-body and $2$-body process,
\begin{align}
dPS_{n+2}=dPS_{n}\times dPS_2(q; p,p')\times \frac{dq^2}{2\pi}\ .
\end{align}
This allows us to write the integrated 
polarization interference $\mathbb{I}_{\rm pol}$,
summed over the helicities of $f,\overline{f'}$, as  
\begin{align}
\label{eq:non_interference_integral}
\frac{d\mathbb{I}_{\rm pol}^{V+\jets}}{dq^2 dPS_{n}}\ 
&=\ \frac{1}{2\pi}\frac{1}{\mathcal{F}}\sum_{h,h'}\int dPS_2\ 
\mathcal{I}_{\rm pol}^{V+\jets}(h,h')
\\
=\ 
\frac{1}{(4\pi)^3}&\frac{1}{\mathcal{F}}\Re\left[\sum_{h,h',k}\int
d\bar\phi_f d\cos\bar\theta_f\ I^{k}_{h,h'} \right]\ .
\end{align}
Here, $\mathcal{F}$ encapsulates the usual flux and 
initial-state averaging factors 
that enter the definition of cross sections.

Due to rotational symmetry in $1\to2$ splitting,
each $I^{k}_{h,h'}$ vanishes 
after integrating over the azimuthal angle,
\begin{align}
\label{eq:non_interference_integral_phi}
\int_0^{2\pi}d\bar\phi_f\ I^{k}_{h,h'}\ =\ 0\ .
\end{align}

After integrating instead over the polar angle, we get
\begin{subequations}
\label{eq:non_interference_integral_theta}
\begin{align}
\int_{-1}^{+1} d\cos\bar\theta_f &\left(I^{1}_{LR} + I^{1}_{RL}\right) 
= 
\nonumber\\
+&\frac{i\pi}{2}\ 
\left(\vert g_L^f\vert^2-\vert g_R^f\vert^2\right)\ 
\tilde{r}_{in}^{k=1}\ \sin\bar\phi_f\ ,
\\
\int_{-1}^{+1} d\cos\bar\theta_f &\left(I^{2}_{LR} + I^{2}_{RL}\right) 
= 
\nonumber\\
-&\frac{i\pi}{2}\
\left(\vert g_L^f\vert^2-\vert g_R^f\vert^2\right)\ 
\tilde{r}_{in}^{k=2}\ \cos\bar\phi_f\ .
\end{align}    
\end{subequations}

These results are remarkable in several regards:

(i) Polarization interference in $V+\jets$
is tied to forward-backward symmetry  
and hence parity violation in the Standard Model. 
For $\gamma^*$, parity is conserved $(g_L^f=g_R^f)$ 
and the $\overline{\theta}_f$-integrated 
interference vanishes,
in agreement with Ref.~\cite{Lam:1978pu}.
For $Z$, parity is partially violated $(g_L^f\neq g_R^f)$,
and the $\overline{\theta}_f$-integrated 
interference partially vanishes.
For $W$, parity violation is maximal,
and the $\overline{\theta}_f$-integrated interference is non-vanishing.

This explains the differences 
in polarization interference reported 
in Refs.~\cite{Denner:2020bcz,Denner:2020eck}
for $WW$ and $WZ$ production.
For transverse-longitudinal interference 
in $\gamma^*/Z$ interference,
the expression is more complicated 
but still vanishes in the parity-conserving limit $(g_R^{Zff}\to g_L^{Zff})$.

(ii) Our approach to using Green's function $G_{in}^\nu$
allows us to work at a fully exclusive level 
with respect to hadronic activity.
While we recover the all-orders results of 
Lam \& Tung for $\gamma^*$~\cite{Lam:1978pu}, 
their methods relied 
on being fully inclusive with respect to hadronic activity.

(iii) The behavior of polarization interference 
holds for both on- and off-shell $V$
since Eq.~\eqref{eq:polint_vjets} is unintegrated 
over the virtuality of $V$.
The dependence on $q^2$ is encoded in $\tilde{r}_{in}^{k}$
and Eq.~\eqref{eq:def_polint_rk}.
Importantly, when $\sqrt{q^2}/E_V\to0$, 
$\mathcal{M}_{\lambda=0}$ and  
$\mathcal{I}_{\rm pol}^{V+\jets}$ both vanish.
This follows from \textit{massless} spin-1 states carrying 
only transverse polarizations.

(iv) Polarization interference as a fraction 
of the unpolarized cross section scales linearly 
with $1/\gamma_V$.

For this last point we note that 
like $\mathbb{I}_{\rm pol}^{V+\jets}$ 
the parton-level cross section for unpolarized $V+\jets$ 
is 
\begin{align}
\label{eq:unpol_xsec_integral}
\frac{d\hat{\sigma}^{V+\jets}}{dq^2 dPS_{n}}\ 
&=\ \frac{1}{2\pi}\frac{1}{\mathcal{F}}\sum_{h,h'}\int dPS_2\ 
\vert\mathcal{M}_{\rm unpol}(h,h')\vert^2
\\
&=\ 
\frac{2}{3(4\pi)^2}\
\frac{q^2}{\mathcal{F}}\
\frac{\left(\vert g_L^f\vert^2+\vert g_R^f\vert^2\right)}{\vert D_V(q)\vert^2}
\nonumber\\
&\times\left[
\vert\tilde{G}_{in}^{\nu=1}\vert^2
+
\vert\tilde{G}_{in}^{\nu=2}\vert^2
+
\frac{q^2}{E_V^2}
\vert\tilde{G}_{in}^{\nu=3}\vert^2
\right]\ .
\label{eq:unpol_xsec}
\end{align}
One can then quantify polarization interference 
as a fraction of the unpolarized partonic cross section 
by the ratio
\begin{align}
\label{eq:interference_ratio}
\mathcal{R}_{\rm pol}^{V+\jets}\ &\equiv\
\left\vert
\frac{d\mathbb{I}_{\rm pol}^{V+\jets}\ /\ dPS_{n+2}}{
d\hat{\sigma}^{V+\jets}\ /\ dPS_{n}}
\right\vert\
\lesssim\ 
\frac{3}{8\pi}
\frac{\sqrt{q^2}}{E_V}\ .
\end{align}

While acceptance and selection cuts 
in realistic LHC analyses 
will disrupt the cancellations in 
Eq.~\eqref{eq:non_interference_integral_phi}
and
Eq.~\eqref{eq:non_interference_integral_theta},
angular factors are at most $\mathcal{O}(1)$,
meaning that polarization interference is at most 
$\mathcal{O}(10\%)$ in $V+\jets$.
Still, one can estimate the impact of fiducial cuts
by using lepton acceptance and selection,
which are typically 
$\mathcal{A}\times\varepsilon\sim 20\%-50\%$
for $p_T^\ell > 25\GeV$ and $\vert\eta^\ell\vert<2.5$,
giving 
\begin{align}
\label{eq:interference_ratio_fiducial}
\int_{\rm fiducial} dPS_{2}\
\mathcal{R}_{\rm pol}^{V+\jets}\ &\lesssim\ 
\frac{3}{8\pi}
\frac{\sqrt{q^2}}{E_V}\ (1-\mathcal{A}\times\varepsilon) .
\end{align}

This relationship is a key takeaway of our work.
Taking $p_T^Z \sim 50\GeV\ (100\GeV)\ [250\GeV]$,
which are realistic choices in $Z+\jets$ analyses at $\sqrt{s}=13\TeV$~\cite{CMS:2022vkb,ATLAS:2024xxl},
$E_Z\approx\sqrt{M_Z^2+(p_T^Z)^2}$,
and approximating $\mathcal{A}\times\varepsilon=50\%$,
we estimate interference in $Z+\jets$ to be
\begin{align}
  \label{eq:polint_numbers}
\int_{\rm fiducial} dPS_{2}\
\mathcal{R}_{\rm pol}^{V+\jets}\ &\lesssim\  
5\%\ (4\%)\ [2\%]\ .
\end{align}
Similar fractions are expected for resonant $W$ production
with smaller fraction expected for lower-mass $\gamma^*$.

\textbf{Electroweak axial gauge:}
As a check of our result's robustness, 
we note that propagator for an unpolarized gauge boson
in the 4-dimensional EW axial gauge
is~\cite{Dams:2004vi}
\begin{align}
\label{eq:prop_unpol_axial}
\Pi_{\mu\nu}^V(q)& 
    {\Big\vert}_{\rm axial} =
  \cfrac{-i}{D_V(q^2)}\nonumber\\
  & \left[g_{\mu\nu} - 
\cfrac{n_\mu q_\nu + n_\nu q_\mu}
{(n\cdot q)} 
+ \cfrac{n^2}{(q\cdot n)^2}
q_\mu q_\nu\right] .
\end{align}
Here, $n^\mu$ is a gauge-fixing device and therefore 
the values it can take on are more restricted than
those allowed in Eq.~\eqref{eq:prop_pol_tensors}~\cite{Capper:1981rd,Leibbrandt:1994wj}.
Setting $q^\mu = (1,-\hat{q})$ is sometimes 
called the ``parton shower gauge''~\cite{Nagy:2007ty,
Nagy:2014mqa,Hagiwara:2020tbx,Chen:2022gxv}.

Since we are working at lowest order in the EW 
theory with massless particles, 
the external currents $G_{in}^\nu$ and $J^\mu_{h,h'}$
are the same as in the $R_\xi$ gauge.
Applying current conservation
leaves only the $g_{\mu\nu}$ term
and we recover the same unpolarized matrix element 
as in the $R_\xi$ gauge,
in accordance with gauge and BRST invariance.

In this gauge, 
the transverse polarization vectors,
and hence the transverse matrix element,
are again the same as in the $R_\xi$ gauge.
The longitudinal propagator is~\cite{Basu:2025zds}
\begin{align}
\label{eq:prop_long_axial}
    \Pi_{\mu\nu}^V(q,\lambda=0)& 
    {\Big\vert}_{\rm axial} =
    \nonumber\\
\cfrac{i}{D_V(q^2)}&\left[
\cfrac{q^2 n^2}{(q\cdot n)^2}
    \Theta_{\mu\nu} 
    +
    \cfrac{q^2}{(q\cdot n)^2}
    n_\mu n_\nu
\right]
.
\end{align}
Noting that $n^2=0$, 
the longitudinal propagator reduces to the 
$n_\mu n_\nu$ term 
and we obtain the same longitudinal matrix element 
as in the $R_\xi$ gauge.

As there is no $\lambda=S$ polarization 
in axial-type gauges, all the matrix elements 
of the $V+\jets$ process for the EW axial gauge 
are the same as in Eq.~\eqref{eq:inter_rx_matrix}.
Subsequently, the scaling of polarization interference 
given in Eq.~\eqref{eq:interference_ratio_fiducial}
is the same in both gauge classes.

\textbf{Multiboson Processes:}
For multiboson and mixed QCD-EW processes,
external currents do not necessarily obey 
the conservation conditions of 
Eq.~\eqref{eq:current_conservation_lep}
and 
Eq.~\eqref{eq:current_conservation_jets}
due to new topologies and sub-processes.
Under the narrow width or pole approximations,
however, 
these ``non-resonant'' contributions can be 
suppressed~\cite{Denner:1999gp,Denner:2005fg,Denner:2020bcz}.
As a result, one still can build expressions 
for polarization interference 
as given in Eq.~\eqref{eq:polint_vjets},
but compounded for each set of unintegrated decay kinematics.
%

\textbf{Heavy Ion Collisions:}
Weak boson production in heavy ion collisions
is a sensitive probe of nuclear parton density 
functions~\cite{Kovarik:2015cma,
Kusina:2016fxy,Eskola:2016oht,AbdulKhalek:2020yuc} 
and intra-nucleus sort-range 
correlations~\cite{nCTEQ:2023cpo,Fuks:2024ctk}.
Assuming that factorization holds for heavy ion collisions~\cite{Collins:1984kg,
Collins:1985ue,Collins:2011zzd}
and since we are working to at $\alpha_s$,
then we expect our results 
to hold  for the $V+\jets$ process 
in $pA$ and $AA$ collisions at the LHC, 
and motivating the use of weak boson polarization 
as a novel probe of nuclear structure and fragmentation.

\textbf{New Physics Scenarios:}
The existence of new, massive, colorless gauge bosons 
has long been hypothesized in extensions of the SM. 
For example:
new $Z'$ gauge bosons in vector portal models 
for dark matter~\cite{Holdom:1985ag,Holdom:1986eq} 
and 
new chiral $Z'$ and $W'$ bosons in neutrino mass 
models~\cite{Pati:1974yy,Mohapatra:1974hk,
Mohapatra:1974gc,Senjanovic:1975rk,Langacker:1980js,Hewett:1988xc}.

In renormalizable versions of these models, 
the couplings of $V'$ to quarks and leptons 
have the form as Eq.~\eqref{eq:def_lep_current}.
Working then at lowest order in $V'$ couplings,
the masslessness of $f,\overline{f'}$ 
and the color-singlet nature of $V'$
ensure that the external currents remain  conserved.
Subsequently, the unpolarized and polarized 
matrix elements for the $V'+\jets$ process
can be organized as in Eq.~\eqref{eq:inter_rx_matrix},
leading to the same scaling for polarization 
interference.

\textbf{Conclusions:}
The polarized propagator formalism is a powerful framework 
for predicting the production and decay of 
intermediate, polarized weak bosons in high-energy collisions.
Importantly, it successfully describes 
polarization in multiboson processes at the LHC.

Furthering this program, we report that 
the polarized weak boson propagators of 
Refs.~\cite{Ballestrero:2017bxn,Ballestrero:2019qoy}
possess analytical decompositions that 
make manifest in covariant gauges
the power counting of mass-over-energy factors in scattering amplitudes 
long known in axial-type gauges.

With this new technology, 
given in Eq.~\eqref{eq:prop_pol_rx_phi_theta}
and 
Eq.~\eqref{eq:prop_pol_tensors},
we computed at an arbitrary order of $\alpha_s$ the longitudinal-transverse 
polarization interference for the process $pp\to V_\lambda+\jets$,
where $V_\lambda$ is an EW gauge boson 
with helicity $\lambda=0,\pm1$, 
and with $V$ decaying to a pair of massless leptons.
We find conditions that lead to 
percent-level or vanishing polarization 
interference
while remaining fully exclusive 
with respect to jet kinematics.

Two important aspects of our work are its simplicity and applicability.
Given the identities in Eq.~\eqref{eq:prop_pol_rx_phi_theta}
(see also a companion work~\cite{Basu:2025zds}), 
our results follow largely from current conservation
and rotational invariance, 
and not the evaluation of any particular diagram.

Our procedure for organizing and 
efficiently evaluating polarized scattering amplitudes
is therefore broadly applicable to other EW processes.
Our work also demonstrates that the polarized propagator formalism
is more than just a ``tool of convenience'' 
for efficient Monte Carlo simulations.
It is a powerful and rigorous approach that 
can possibly push the state-of-the-art understanding 
of EW boson production at ultra high energies.

\section*{Acknowledgments}
The authors thank
A.\ Denner,
B.\ Fuks,
A.\ Maier,
A.\ Maas,
G.\ Pelliccioli,
S.\ Pl\"atzer,
R.\ Poncelet,
and
K.\ Potamianos
for helpful insights and discussions.

The authors acknowledges the support 
of the Narodowe Centrum Nauki (NCN) 
under Grant No. 2023/49/ B/ST2/04330 (SNAIL). 
The authors acknowledge support from the COMETA COST Action CA22130.



\bibliography{polarInt_Vjets_refs.bib}

\end{document}